\documentclass[twocolumn,aps,showpacs,prl,superscriptaddress,floatfix]{revtex4}
\usepackage{graphicx}
\usepackage{amsmath,amssymb}
\usepackage{bm}

\input{tcilatex}

\begin{document}

\title{Single-atom as a macroscopic entanglement source}
\author{Ling Zhou}
\affiliation{Institute for Quantum Studies and Department of Physics, Texas A\&M
University, College Station, Texas 77843-4242}
\affiliation{Department of Physics, Dalian University of Technology, Dalian 116024, China}
\author{Han Xiong}
\affiliation{Institute for Quantum Studies and Department of Physics, Texas A\&M
University, College Station, Texas 77843-4242}
\author{M. Suhail Zubairy}
\affiliation{Institute for Quantum Studies and Department of Physics, Texas A\&M
University, College Station, Texas 77843-4242}

\begin{abstract}
We discuss the generation of a macroscopic entangled state in a single atom
cavity-QED system. The three-level atom in a cascade configuration interacts
dispersively with two classical coherent fields inside a doubly resonant
cavity. We show that a macroscopic entangled state between these two cavity
modes can be generated under large detuning conditions. The entanglement
persists even under the presence of cavity losses.
\end{abstract}

\maketitle

\section{Introduction}

Quantum entanglement lies at the heart of quantum computing and quantum
information science. Cavity quantum electrodynamics (QED) provides an
important testing ground for these ideas. For example, cavity QED can be
used to not only store quantum information but also act as a source of
entanglement [1-8]. The generation of entanglement in cavity QED has been
studied by many authors including the generation of entangled coherent state %
\cite{davidovich,zou,solano}, single photon and vacuum entanglement \cite%
{zubairy}, and two-atom entanglement \cite{zheng2}.

More recently, generation of macroscopic entangled states via phase
sensitive amplification has been discussed. Such continuous variables
entanglement offers many advantages in quantum information processing\cite%
{sl}. For example, a quantum secure communication protocol using continuous
variables Einstein-Podolsky-Rosen correlations was proposed in \cite{he}.
Conventionally, continuous variables entanglement is produced in a
parametric down-conversion process \cite{pan}. Recently, based on the study
concerning a two-mode correlated spontaneous emission laser (CEL) \cite%
{scully}, it was shown that a CEL can lead to two-mode entanglement even
when the average photon number can be very large \cite{han,tan}. The scheme
using CEL is the result of many-atom dynamics. The scheme \cite{guz} with
potential to produce macroscopic entangled states is still for atomic cloud.
On the other hand, a one-atom laser, has been realized experimentally \cite%
{kimble}. The entanglement between single atom and its emitted photon has
been observed \cite{bl}. More recently, Morigi et al \cite{gm, gm2} put
forward a scheme where a single trapped atom allows for the generation of
entangled light under certain conditions.

In this paper, we propose a scheme to produce a macroscopic entangled state
using a single atom in a cavity QED system. We show that a two-mode coherent
squeezed state can be generated from our system. In our scheme, a driven
three-level atom in cascade configuration dispersively interacts with a
two-mode field. We show that under appropriate conditions on the detunings
and atom-field coupling, the classical driving fields can help to build up
the field in the two modes of the cavity and at the same time an
entanglement is generated between the two modes.

\section{System description and calculations}

We consider a three-level atom in a cascade configuration crossing or
trapped in a two-mode field cavity. The atomic level configuration is
depicted in Fig. (\ref{fig1}). The two atomic transitions $|a\rangle
\leftrightarrow |b\rangle $ and $|b\rangle \leftrightarrow |c\rangle $
interact with the two cavity modes with detunings $\mp \delta $ with $\delta
=|\omega _{1}-(E_{a}-E_{b})|=|\omega _{2}-(E_{b}-E_{c})|$. The two atomic
transitions (namely, $|a\rangle \leftrightarrow |b\rangle $ and $|b\rangle
\leftrightarrow |c\rangle $) are also driven by two classical fields with
the same detunings as their corresponding quantized field modes and $\Omega
_{1}$ and $\Omega _{2}$ are the Rabi frequencies of the two classical
fields. The dipole forbidden atomic transition between $|a\rangle $ and $%
|c\rangle $ are resonantly driven by another classical field of Rabi
frequency $\Omega $.\FRAME{ftbpFU}{2.5425in}{2.5123in}{0pt}{\Qcb{The level
configuration of three-level atom. Two cavity modes and two classical fields
interact with atomic transitions $|a\rangle \leftrightarrow |b\rangle $ and $%
|b\rangle \leftrightarrow |c\rangle $ with the detunings $\mp \protect\delta 
$, and another classical field with the Rabi frequency of $\Omega $ drives
the dipole forbidden atomic transition between $|a\rangle $ and $|c\rangle $
resonantly.}}{}{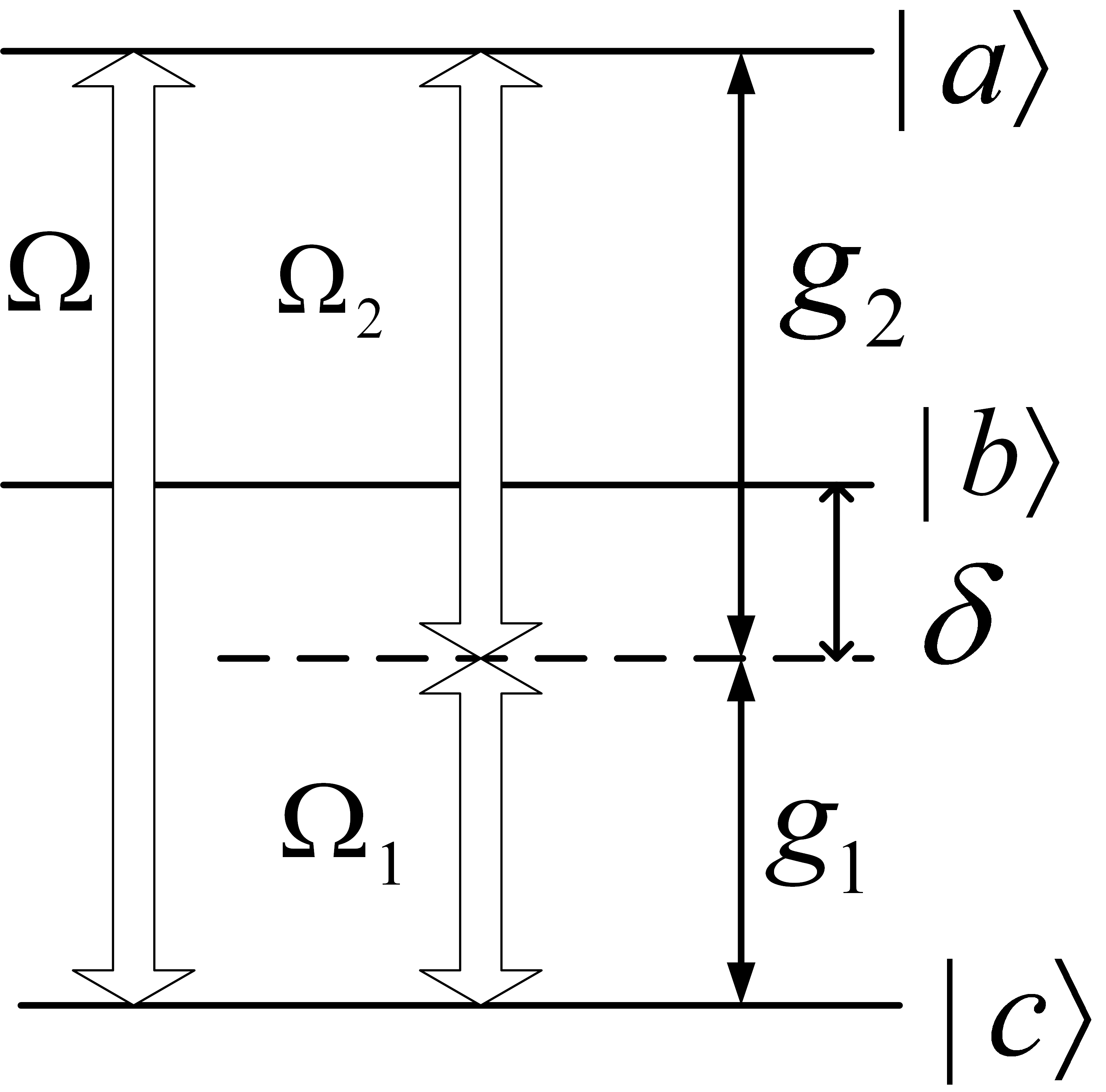}{\special{language "Scientific Word";type
"GRAPHIC";maintain-aspect-ratio TRUE;display "USEDEF";valid_file "F";width
2.5425in;height 2.5123in;depth 0pt;original-width 31.3754in;original-height
30.9897in;cropleft "0";croptop "1";cropright "1";cropbottom "0";filename
'fig1.jpg';file-properties "XNPEU";}} 

The Hamiltonian of our system under the dipole and rotating wave
approximation and in the interaction picture is given by 
\begin{eqnarray}
\hat{H}_{I} &=&g_{1}[(\hat{a}_{1}+\frac{\Omega _{1}}{g_{1}})\hat{\sigma}%
_{bc}+(\hat{a}_{1}^{\dagger }+\frac{\Omega _{1}}{g_{1}})\hat{\sigma}_{cb}] 
\notag  \label{eq1} \\
&&+g_{2}[(\hat{a}_{2}+\frac{\Omega _{2}}{g_{2}})\hat{\sigma}_{ab}+(\hat{a}%
_{2}^{\dagger }+\frac{\Omega _{2}}{g_{2}})\hat{\sigma}_{ba}]  \notag \\
&&+\Omega (\hat{\sigma}_{ac}+\hat{\sigma}_{ca})-\delta (\hat{\sigma}_{aa}+%
\hat{\sigma}_{cc})
\end{eqnarray}%
where $\hat{\sigma}_{ij}=|i\rangle \langle j|$ ($i,j=a,b,c$) are the atomic
operators. $\hat{a}_{1}(\hat{a}_{1}^{\dagger })$ and $\hat{a}_{2}(\hat{a}%
_{2}^{\dagger })$ are the creation (annihilation) operators of the two
cavity modes and $g_{1}$ and $g_{2}$ are the atom-field coupling constants
and in general they are different.

The Heisenberg equations of motion for the atomic operators $\hat{\sigma}%
_{bc}$ and $\hat{\sigma}_{ba}$ are given by 
\begin{equation}
\left\{ 
\begin{array}{c}
i\frac{d\hat{\sigma}_{bc}}{dt}=-g_{1}\tilde{\hat{a}}_{1}^{\dagger }(\hat{%
\sigma}_{cc}-\hat{\sigma}_{bb})-g_{2}\tilde{\hat{a}}_{2}\sigma _{ac} \\ 
+\Omega \hat{\sigma}_{ba}-\delta \hat{\sigma}_{bc}, \\ 
i\frac{d\hat{\sigma}_{ba}}{dt}=-g_{1}\tilde{\hat{a}}_{1}^{\dagger }\hat{%
\sigma}_{ca}+g_{2}\tilde{\hat{a}}_{2}(\hat{\sigma}_{bb}-\hat{\sigma}_{aa})
\\ 
+\Omega \hat{\sigma}_{bc}-\delta \hat{\sigma}_{ba},%
\end{array}%
\right.  \label{eq2}
\end{equation}%
where%
\begin{equation}
\tilde{a}_{j}=\hat{a}_{j}+\frac{\Omega _{j}}{g_{j}},\tilde{a}_{j}^{\dagger }=%
\hat{a}_{j}^{\dagger }+\frac{\Omega _{j}}{g_{j}},j=1,2.
\end{equation}%
Under the large detuning condition when $\delta \gg \Omega ,\Omega
_{j},g_{1},g_{2}$, Eq. (\ref{eq2}) can be solved adiabatically by taking ${d%
\hat{\sigma}_{bc}}/{dt}={d\hat{\sigma}_{ba}}/{dt}=0$. The adiabatic
solutions for $\hat{\sigma}_{bc}$ and $\hat{\sigma}_{ab}$ can then be
substituted into the Hamiltonian (\ref{eq1}) and we obtain 
\begin{eqnarray}
\hat{H}_{1} &=&\Omega (\hat{\sigma}_{ac}+\hat{\sigma}_{ca})-\delta (\hat{%
\sigma}_{aa}+\hat{\sigma}_{cc})  \notag  \label{eq3} \\
&&+\frac{1}{\Omega ^{2}-\delta ^{2}}\{-\delta g_{1}^{2}(2\tilde{a}%
_{1}^{\dagger }\tilde{a}_{1}+1)(\hat{\sigma}_{bb}-\hat{\sigma}_{cc})  \notag
\\
&&+\delta g_{2}^{2}(2\tilde{a}_{2}^{\dagger }\tilde{a}_{2}+1)(\hat{\sigma}%
_{aa}-\hat{\sigma}_{bb})  \notag \\
&&+\Omega \lbrack (g_{1}^{2}\tilde{a}_{1}^{\dagger }\tilde{a}_{1}+g_{2}^{2}%
\tilde{a}_{2}^{\dagger }\tilde{a}_{2})\hat{\sigma}_{ac}  \notag \\
&&+(g_{1}^{2}\tilde{a}_{1}\tilde{a}_{1}^{\dagger }+g_{2}^{2}\tilde{a}_{2}%
\tilde{a}_{2}^{\dagger })\hat{\sigma}_{ca}] \\
&&+2g_{1}g_{2}\delta (\tilde{a}_{1}\tilde{a}_{2}\hat{\sigma}_{ac}+\tilde{a}%
_{1}^{\dagger }\tilde{a}_{2}^{\dagger }\hat{\sigma}_{ca})  \notag \\
&&+g_{1}g_{2}\Omega (\tilde{a}_{1}\tilde{a}_{2}+\tilde{a}_{1}^{\dagger }%
\tilde{a}_{2}^{\dagger })(\hat{\sigma}_{aa}+\hat{\sigma}_{cc}-2\hat{\sigma}%
_{bb})\}.  \notag
\end{eqnarray}%
If the atom is initially injected in level $|b\rangle $, it will remain
confined to this level due to the large detuning approximation. The
approximate effective Hamiltonian for this case reduces to 
\begin{equation}
\hat{H}_{b}=\eta _{1}\tilde{a}_{1}^{\dagger }\tilde{a}_{1}+\eta _{2}\tilde{a}%
_{2}^{\dagger }\tilde{a}_{2}+\frac{1}{2}(\eta _{1}+\eta _{2})+\xi (\tilde{a}%
_{1}\tilde{a}_{2}+\tilde{a}_{1}^{\dagger }\tilde{a}_{2}^{\dagger })
\label{5}
\end{equation}%
where 
\begin{eqnarray}
\xi &=&\frac{2g_{1}g_{2}\Omega }{\delta ^{2}-\Omega ^{2}},  \notag \\
\eta _{1} &=&\frac{2g_{1}^{2}\delta }{\delta ^{2}-\Omega ^{2}},  \label{6} \\
\eta _{2} &=&\frac{2g_{2}^{2}\delta }{\delta ^{2}-\Omega ^{2}}.  \notag
\end{eqnarray}%
This Hamiltonian can be rewritten as 
\begin{equation}
\hat{H}_{b}=(\eta _{1}+\eta _{2})\hat{K}_{0}+\xi (\hat{K}_{-}+\hat{K}_{+})+%
\frac{1}{2}(\eta _{1}-\eta _{2})\hat{N}_{0},  \label{7}
\end{equation}%
where 
\begin{eqnarray*}
\hat{K}_{0} &=&\frac{1}{2}(\tilde{a}_{1}^{\dagger }\tilde{a}_{1}+\tilde{a}%
_{2}^{\dagger }\tilde{a}_{2}+1), \\
\hat{K}_{-} &=&\tilde{a}_{1}\tilde{a}_{2}, \\
\hat{K}_{+} &=&\tilde{a}_{1}^{\dagger }\tilde{a}_{2}^{\dagger }, \\
\hat{N}_{0} &=&\tilde{a}_{1}^{\dagger }\tilde{a}_{1}-\tilde{a}_{2}^{\dagger }%
\tilde{a}_{2}.
\end{eqnarray*}%
These operators can be verified to obey the $SU(1,1)$ commutation relations $%
[\hat{K}_{-},\hat{K}_{+}]=2\hat{K}_{0}$, $[\hat{K}_{0},\hat{K}_{\pm }]=\pm 
\hat{K}_{\pm }$, and $[\hat{N}_{0},\hat{K}_{0}]=[\hat{N}_{0,}\hat{K}_{\pm
}]=0$. We can therefore use the $SU(1,1)$ Lie-algbra to expand the unitary
evolution \cite{BAN} operator $\hat{U}=e^{-i\hat{H}_{b}t}$ as 
\begin{equation}
\hat{U}=e^{(A_{+}\hat{K}_{+})}e^{(\ln A_{0}\hat{K}_{0})}e^{-\frac{it}{2}%
(\eta _{1}-\eta _{2})\hat{N}_{0}}e^{(A_{-}\hat{K}_{-})}  \label{8}
\end{equation}%
where%
\begin{eqnarray}
A_{0} &=&a_{0}^{2},  \notag \\
A_{+} &=&A_{-}=\frac{-i\xi t}{\phi }a_{0}\sinh \phi  \label{9}
\end{eqnarray}%
with%
\begin{eqnarray}
a_{0} &=&\frac{1}{\cosh \phi +it\frac{(\eta _{1}+\eta _{2})}{2\phi }\sinh
\phi },  \notag \\
\phi ^{2} &=&[-(\frac{\eta _{1}+\eta _{2}}{2})^{2}+\xi ^{2}]t^{2}.
\label{10}
\end{eqnarray}

We now consider the case when the two-mode field is initially prepared in a
vacuum state $|0,0\rangle $. The time evolution of the field state can be
obtained as 
\begin{equation}
|\Psi _{f}(t)\rangle \Rightarrow \exp (A_{+}\hat{a}_{1}^{\dagger }\hat{a}%
_{2}^{\dagger })\exp (\alpha _{1}\hat{a}_{1}^{\dagger })\exp (\alpha _{2}%
\hat{a}_{2}^{\dagger })|0,0\rangle
\end{equation}%
with 
\begin{eqnarray}
\alpha _{1} &=&\frac{\Omega _{2}}{g_{2}}A_{+}+\frac{\Omega _{1}}{g_{1}}%
[a_{0}e^{-\frac{it}{2}(\eta _{1}-\eta _{2})}-1],  \notag  \label{eq9} \\
\alpha _{2} &=&\frac{\Omega _{1}}{g_{1}}A_{+}+\frac{\Omega _{2}}{g_{2}}%
[a_{0}e^{\frac{it}{2}(\eta _{1}-\eta _{2})}-1].
\end{eqnarray}%
The $SU(1,1)$ Lie-algebra yields 
\begin{equation*}
e^{A_{+}\hat{a}_{1}^{\dagger }\hat{a}_{2}^{\dagger }}=e^{(\vartheta ^{\ast }%
\hat{a}_{1}\hat{a}_{2}-\vartheta \hat{a}_{1}^{\dagger }\hat{a}_{2}^{\dagger
})}e^{A_{+}^{\ast }\hat{a}_{1}\hat{a}_{2}}e^{g(\hat{a}_{1}^{\dagger }\hat{a}%
_{1}+\hat{a}_{2}^{\dagger }\hat{a}_{2}+1)}.
\end{equation*}%
Let $\vartheta =re^{i\varepsilon }$, $g=\ln \cosh r$ \ where $r$ and $%
\varepsilon $ are determined by the relation 
\begin{equation}
A_{+}=-e^{i\varepsilon }\tanh r.  \label{13}
\end{equation}%
The squeezed parameter $r$ and $\varepsilon $ are

\begin{eqnarray}
r &=&\tanh ^{-1}|A_{+}|, \\
\cos \varepsilon &=&-\frac{\func{Re}(A_{+})}{|A_{+}|},  \notag \\
\sin \varepsilon &=&-\frac{\func{Im}(A_{+})}{|A_{+}|}.  \notag
\end{eqnarray}%
The state of the system can then be written as 
\begin{eqnarray}
|\Psi _{f}(t)\rangle &=&e^{(\vartheta ^{\ast }\hat{a}_{1}\hat{a}%
_{2}-\vartheta \hat{a}_{1}^{\dagger }\hat{a}_{2}^{\dagger })}|\alpha
_{1}\cosh r,\alpha _{2}\cosh r\rangle  \notag  \label{eq10} \\
&=&S(\vartheta )D(\alpha _{1}\cosh r)D(\alpha _{2}\cosh r)|0,0\rangle .
\end{eqnarray}%
It is obviously a two-mode coherent-squeezed state \cite{zeng}\cite{lee}.

For the generation of macroscopic entangled state, we consider two
quantities, namely the mean photon number and the correlation functions
involved in the entanglement criterion. The total average photon number of
the two-mode field $N=\langle \hat{a}_{1}^{\dagger }\hat{a}_{1}\rangle
+\langle \hat{a}_{2}^{\dagger }\hat{a}_{2}\rangle $ can be easily obtained 
\begin{eqnarray}
N &=&2\sinh ^{2}r+\cosh ^{2}r[(|\alpha _{1}|^{2}+|\alpha _{2}|^{2})\cosh 2r 
\notag \\
&&-(\alpha _{1}\alpha _{2}e^{-i\varepsilon }+\alpha _{1}^{\ast }\alpha
_{2}^{\ast }e^{i\varepsilon })\sinh 2r].  \label{16}
\end{eqnarray}%
To determine the entanglement of state (\ref{eq10}), we need the
entanglement criterion for continuous variables system. Recently, different
criteria have been proposed \cite{duan,vit, shchu,zubairy2}. Here, we choose
the summation of the quantum fluctuations proposed in Ref. \cite{duan}.
According to this criterion, a state is entangled if the summation of the
quantum fluctuations in the two EPR-like operators $\hat{u}$ and $\hat{v}$
satisfy the following inequality 
\begin{equation}
(\Delta \hat{u})^{2}+(\Delta \hat{v})^{2}<2,  \label{eq17}
\end{equation}%
where 
\begin{equation*}
\hat{u}=\hat{x}_{1}+\hat{x}_{2,}\hat{v}=\hat{p}_{1}-\hat{p}_{2},
\end{equation*}%
and $\hat{x}_{i}=(\hat{a}_{i}e^{-i\psi }+\hat{a}_{i}^{\dagger }e^{i\psi })/%
\sqrt{2}$ and $\hat{p}_{i}=(\hat{a}_{i}e^{-i\psi }-\hat{a}_{i}^{\dagger
}e^{i\psi })/\sqrt{2}i$ ($i=1,2$) are the quadrature operators of the field.
For the state (15) and by taking $\psi =\frac{1}{4}\pi $ we can derive that 
\begin{equation}
(\Delta \hat{u})^{2}+(\Delta \hat{v})^{2}=2(\cosh 2r-\sin \varepsilon \sinh
2r).  \label{eq18}
\end{equation}%
From Eqs. (12), (16) and (18), it is clear that the average photon number of
the two-mode field depends on $\Omega _{1}$ and $\Omega _{2}$, however, the
entanglement condition is independent of the strengths of the driving
fields. We can thus change the average photon number of the field by
manipulating $\Omega _{1}$ and $\Omega _{2}$ without affecting the
entanglement of the two modes.

It is useful to consider the case when $\Omega _{1}=\Omega _{2}=0$ which
means $\alpha _{1}=\alpha _{2}=0$ in Eq. (12). From Eq. (15), we have 
\begin{eqnarray}
|\Psi \rangle &=&S(\vartheta )|0,0\rangle  \notag \\
&=&\frac{1}{\cosh r}\sum_{n}\tanh ^{n}r|n,n\rangle  \label{eq19}
\end{eqnarray}%
This is a two-mode squeezed state which can also be generated by a
parametric amplifier \cite{wall}. The total average photon number of the
two-mode field for an initial vacuum state is 
\begin{equation}
N=2\sinh ^{2}r.  \label{eq20}
\end{equation}%
The entanglement condition still has the form of Eq.(18).

Next we consider the effect of the cavity losses by including the cavity
damping terms in the equation of motion for the density operators. The
equation of motion for the density operator is given by 
\begin{eqnarray}
\dot{\hat{\rho}} &=&-i[\xi (\hat{a}_{1}\hat{a}_{2}+\hat{a}_{1}^{\dagger }%
\hat{a}_{2}^{\dagger })+\eta _{1}\hat{a}_{1}^{\dagger }\hat{a}_{1}+\eta _{2}%
\hat{a}_{2}^{\dagger }\hat{a}_{2}  \notag  \label{eq15} \\
&&+(\eta _{2}\frac{\Omega _{2}}{g_{2}}+\xi \frac{\Omega _{1}}{g_{1}})(\hat{a}%
_{2}+\hat{a}_{2}^{\dagger })  \notag \\
&&+(\eta _{1}\frac{\Omega _{1}}{g_{1}}+\xi \frac{\Omega _{2}}{g_{2}})(\hat{a}%
_{1}+\hat{a}_{1}^{\dagger }),\hat{\rho}]  \notag \\
&&+\kappa \sum_{i=1,2}(2\hat{a}_{i}\hat{\rho}\hat{a}_{i}^{\dagger }-\hat{a}%
_{i}^{\dagger }\hat{a}_{i}\hat{\rho}-\hat{\rho}\hat{a}_{i}^{\dagger }\hat{a}%
_{i}).
\end{eqnarray}%
The resulting equations for the expectation values of the field operators
are 
\begin{eqnarray}
\frac{d\langle \hat{a}_{1}^{\dagger }\hat{a}_{1}\rangle }{dt} &=&-i[\xi
(\langle \hat{a}_{1}^{\dagger }\hat{a}_{2}^{\dagger }\rangle -\langle \hat{a}%
_{1}\hat{a}_{2}\rangle )  \notag \\
&&+(\eta _{1}\frac{\Omega _{1}}{g_{1}}+\xi \frac{\Omega _{2}}{g_{2}}%
)(\langle \hat{a}_{1}^{\dagger }\rangle -\langle \hat{a}_{1}\rangle
)]-2\kappa \langle \hat{a}_{1}^{\dagger }\hat{a}_{1}\rangle ,  \notag \\
\frac{d\langle \hat{a}_{1}\hat{a}_{2}\rangle }{dt} &=&-i[\xi (\langle \hat{a}%
_{1}^{\dagger }\hat{a}_{1}\rangle +\langle \hat{a}_{2}^{\dagger }\hat{a}%
_{2}\rangle +1)  \notag \\
&&+(\eta _{1}\frac{\Omega _{1}}{g_{1}}+\xi \frac{\Omega _{2}}{g_{2}})\langle 
\hat{a}_{2}\rangle +(\eta _{2}\frac{\Omega _{2}}{g2}+\xi \frac{\Omega _{1}}{%
g_{1}})\langle \hat{a}_{1}\rangle ]  \notag \\
&&-[2\kappa +i(\eta _{1}+\eta _{2})]\langle \hat{a}_{1}\hat{a}_{2}\rangle ,
\label{eq22} \\
\frac{d\langle \hat{a}_{1}\rangle }{dt} &=&-i[\xi \langle \hat{a}%
_{2}^{\dagger }\rangle +(\eta _{1}\frac{\Omega _{1}}{g_{1}}+\xi \frac{\Omega
_{2}}{g_{2}})]-(\kappa +i\eta _{1})\langle \hat{a}_{1}\rangle .  \notag
\end{eqnarray}%
On interchanging the subscripts 1 and 2 and taking the Hermitian conjugate,
we can obtain the remaining five differential equations of $\langle \hat{a}%
_{2}^{\dagger }\hat{a}_{2}\rangle $, $\langle \hat{a}_{1}^{\dagger }\hat{a}%
_{2}^{\dagger }\rangle $ ect.. These eight equations can be solved by using
the standard techniques such as those based on Laplace transform method. We
can then evaluate the average photon numbers and the quantity $(\Delta 
\hat{u})^{2}+(\Delta \hat{v})^{2}$ for this system. These solutions are long
and tedious and we do not reproduce them here. Instead, we present a
numerical solutions for these equations in the next section.

\section{Discussion}

\begin{figure}[h]
\includegraphics*[width=80mm,height=50mm]{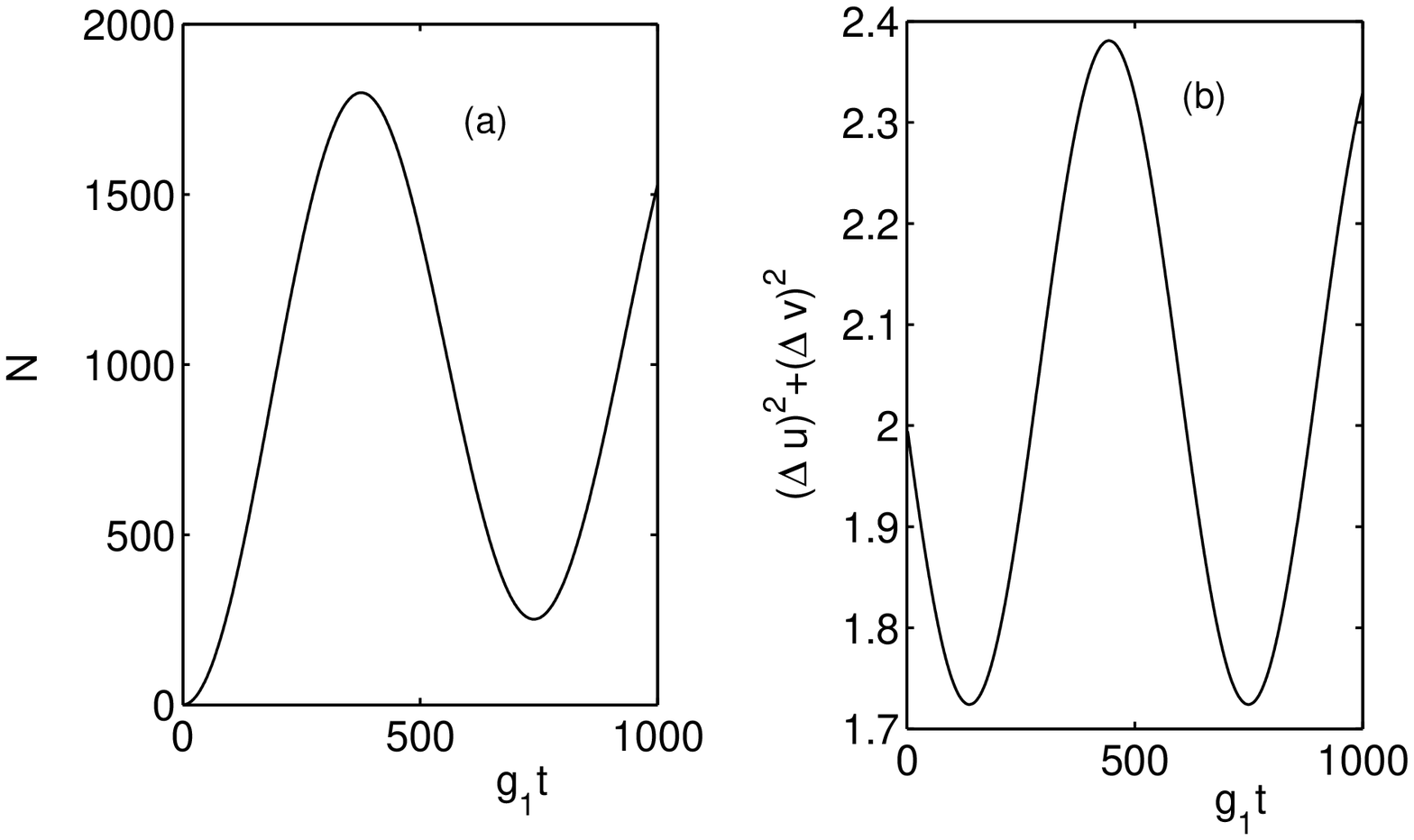}
\caption{The time evolution of the total average photon number $N$ \ and $%
(\Delta \hat{u})^{2}+(\Delta \hat{v})^{2}$. The two-mode field is entangled
when $(\Delta \hat{u})^{2}+(\Delta \hat{v})^{2}<2$ (Eq. (15)) . The
parameters are $\Omega _{1}=10$, $\Omega _{2}=40$, $g_{1}=1$, $g_{2}=2$, $%
\protect\delta =1000$, and $\Omega =200$. }
\label{fig2}
\end{figure}

\begin{figure}[h]
\includegraphics*[width=85mm,height=50mm]{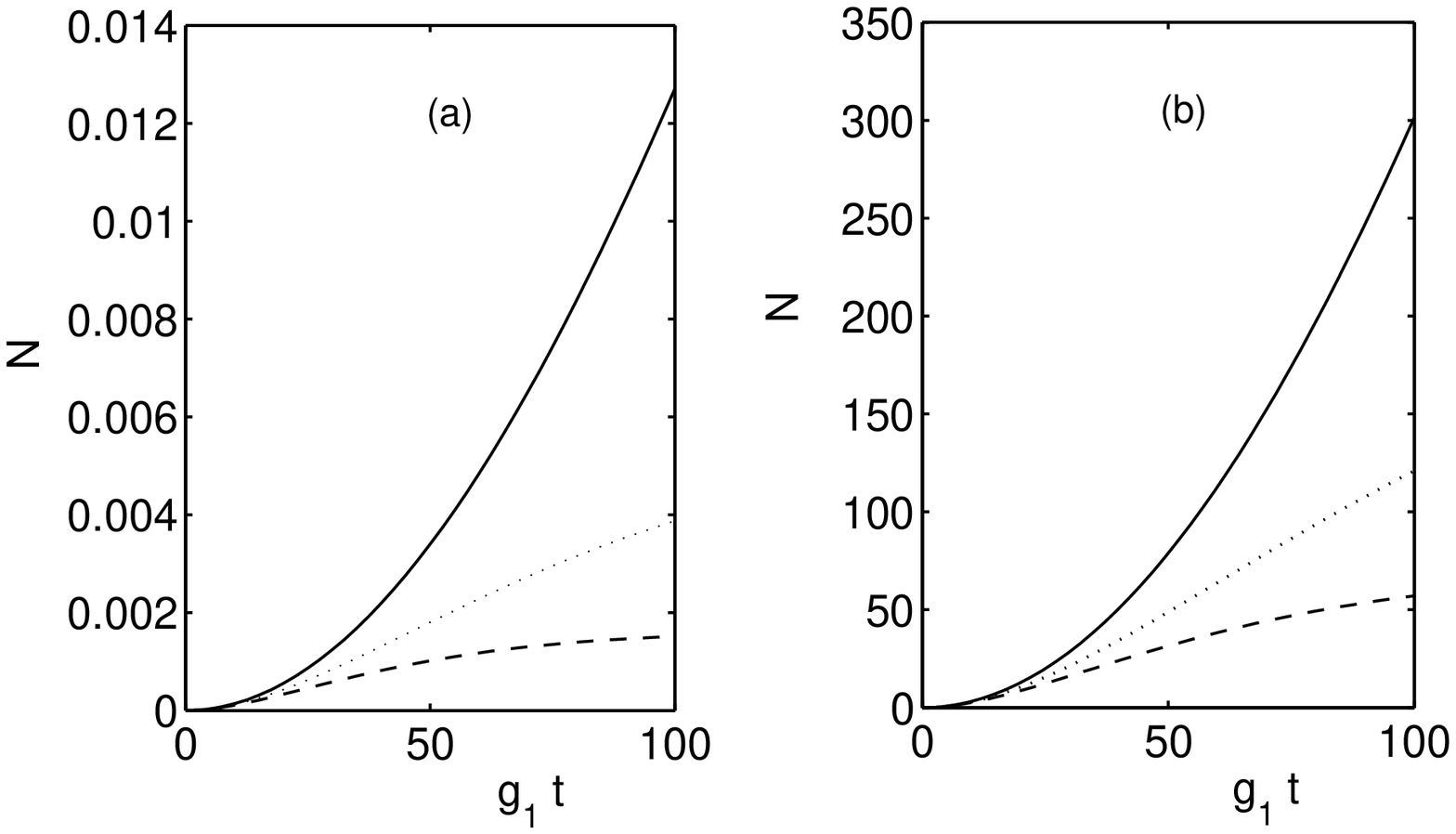}
\caption{The time evolution of total average photon number $N$ . Solid lines
in 2a and 2b correspond to two-mode squeezed vacuum state (Eq.(19)) and the
two-mode coherent-squeezed state (Eq.(16)), respectively; dotted ( $\protect%
\kappa =0.01$) and dashed lines ($\protect\kappa =$ $0.02)$ are plotted from
Eq. (21). In Fig. 2a, $\Omega _{1}=\Omega _{2}=0$ while for Fig. 2b $\Omega
_{1}=10$, $\Omega _{2}=40.$ For all plots, $g_{1}=1$, $g_{2}=2$ and $\protect%
\delta =1000$, $\Omega =200$. }
\label{fig3}
\end{figure}

We now discuss the entanglement properties of the amplified fields inside
the doubly resonant cavity. In our plots, all of parameters are in expressed
in units of $g_{1}$. In Fig.2, we plot the average photon number and $%
(\Delta \hat{u})^{2}+(\Delta \hat{v})^{2}$ for the two-mode
coherent-squeezed state from Eqs.(16) and (18), respectively. Both the
quantities exhibit oscillations. This is a consequence of the terms
prportional to $\eta _{1}$ and $\eta _{2}$ in the Hamiltonian (5). The
period of the oscillations can however be very large as $\eta _{1}$ and $%
\eta _{2}$ can be small. Thus we can have entanglement for a sufficiently
large interaction times.

In Figs. 3 and 4, we plot $N$ and $(\Delta \hat{u})^{2}+(\Delta \hat{v})^{2}$
in the small time region where entanglement is present. In Fig. 3 we plot
the total average photon number $N$ as a function of time under two cases: $%
\Omega _{1}=\Omega _{2}=0$ (Fig. 2a) and $\Omega _{1}=10$, $\Omega _{2}=40$
(Fig. 2b). \ Solid lines in Figs. 2a and 2b are plotted from Eqs. (20) and
(16), respectively. Dotted lines and dashed lines are plotted from Eq. (21)
with the inclusion of cavity losses. Comparing the two solid lines in Fig.
2a and 2b, we note that the average photon number of two-mode
coherent-squeezed state is extremely larger as compared to a two-mode
squeezed vacuum state. Even with the inclusion of cavity losses (dotted
lines and dashed lines), the average photon number of two-mode fields still
increase dramatically for the driven system. Thus the two-mode fields still
can be amplified even when cavity losses are present.

\begin{figure}[tbph]
\includegraphics*[width=65mm,height=50mm]{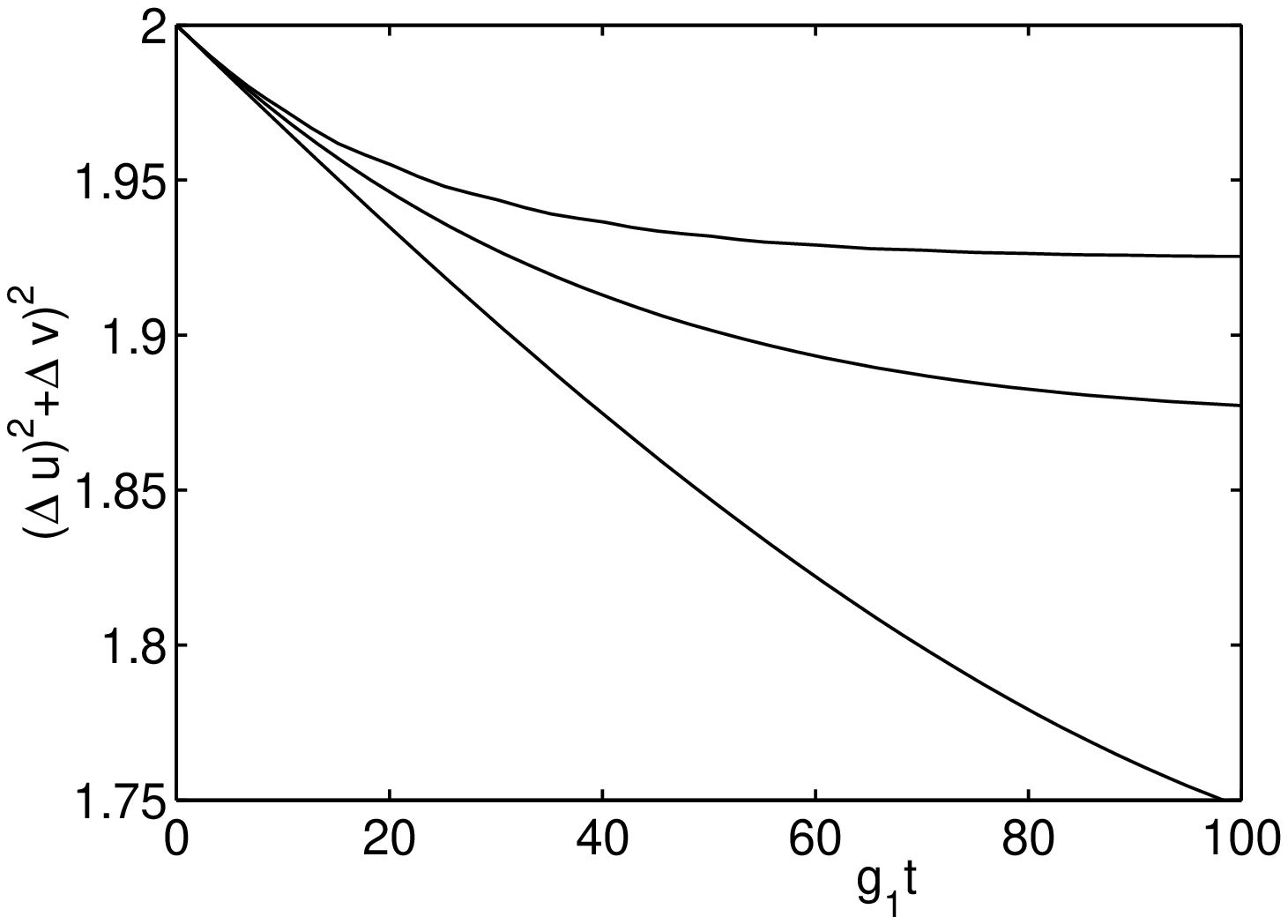}
\caption{The quantity $(\Delta \hat{u})^{2}+(\Delta \hat{v})^{2}$ changes
with time for several values of the cavity decay rates $\protect\kappa $.
From top to bottom $\protect\kappa =0.02,0.01,0$. Other parameters are $%
\Omega =200,g_{1}=1$, $g_{2}=2$, and $\protect\delta =1000$.}
\label{fig4}
\end{figure}

In Fig. 4, we show the time evolution of $(\Delta \hat{u})^{2}+(\Delta \hat{v%
})^{2}$ in the presence of cavity losses. Notice that the entanglement
exists in a lossy cavity. It is worthwhile to point out that we plot $%
(\Delta \hat{u})^{2}+(\Delta \hat{v})^{2}$ and $N$ for the same set of
parameters. For this set of parameters, we obtain both amplification and
entanglement at the same time.

Finally, we note that the classical field $\Omega $ can not only affect
entanglement between the the two quantum fields but also the amplification
of these fields. On the other hand, the two classical fields $\Omega _{1%
\text{ }}$and $\Omega _{2}$ mainly amplify the quantum field but plays no
role on the entanglement criterion.

We note that Morigi et al \cite{gm,gm2} have also considered the generation
of two-mode squeezing in a single atom. Their situation is however different
from ours. In their work, Morigi et al consider both external and internal
degrees of freedom. Under the large detuning limit, the atom's internal
degrees of freedom are eliminated and a two-mode squeezed state at certain
times is obtained. At those times the atom is decorrelated from the two
cavity modes. At other times, the system is in a tripartite entangled state
between the cavity modes and the center-of-mass degrees of freedom of the
atom. In our scheme, the entangled states are generated over a wide range of
interaction times. This is easily seen from Figs. 2 and 4. Moreover in our
scheme, we can generate a two-mode coherent-squeezed state of large
intensity.

\section{Conclusion}

In summary, we discussed a scheme in which a single atom in the cascade
configuration inside a doubly resonant cavity can lead to amplified fields
that are entangled. The resulting field, under appropriate conditions, is a
two-mode coherent-squeeze state. We show that the entanglement persists even
in the presence of cavity losses.

\acknowledgements This work is supported by DARPA-QuIST, the Air Force
Office of Scientific Research and the Office of Naval Research.

\end{document}